\newfam\msbfam
\font\twlmsb=msbm10 at 12pt
\font\eightmsb=msbm10 at 8pt
\font\sixmsb=msbm10 at 6pt
\textfont\msbfam=\twlmsb
\scriptfont\msbfam=\eightmsb
\scriptscriptfont\msbfam=\sixmsb
\def\cj{\fam\msbfam}

\def\I{{\cj I}}

\centerline{\bf QUANTUM MECHANICS AND LEGGETT'S INEQUALITIES}

\

\centerline{\bf M. Socolovsky$^{(*)}$}

\centerline{\it  Instituto de Ciencias Nucleares, Universidad Nacional Aut\'onoma de M\'exico}
\centerline{\it Circuito Exterior, Ciudad Universitaria, 04510, M\'exico D. F., M\'exico} 
\centerline{\it and}
\centerline{\it Instituto de Astronom\'\i a y F\'\i sica del Espacio, Universidad de Buenos Aires-CONICET}
\centerline{\it C1428ZAA, Bs. As., Argentina}

\

{\it We show that when the proper description of the behaviour of individual photons or spin ${{1}\over{2}}$ particles in a spherically symmetric entangled pair is done through the use of the density matrix, the  Leggett's inequality is not violated by quantum mechanics.}

\

{ Key words}: quantum mechanics; non local realism; inequalities; density matrix.

\

{\bf 1. Introduction}

\

Leggett in 2003 $^1$, and Gr\"oblacher {\it et al} in 2007 $^2$, have derived, in the context of non local realism, new Bell's type inequalities which should be contrasted with the predictions of quantum mechanics. The inequalities are for pairs of polarized photons in a mixture state of subensambles, each consisting of an assumed coherent superposition of states with definite polarization for each photon. The claim of this authors is that quantum mechanics violates the inequalities, thus providing a strong argument against the feasibility of non local hidden variable theories (NLHVT's). 

\

However, even if at the ``semiclassical'' level of a hidden variable theory, with or without the assumption of locality, one assumes that each photon of an entangled pair is in a definite state of polarization (represented by unit vectors $\vec{u}$ or $\vec{v}$ in certain directions), we know that in quantum mechanics each subsystem of a given subensamble is not in a pure state, but in a mixed state, so that the correct treatment of it, in particular for the calculation of expectation values, has to be based on the {\it density matrix approach}; in particular one must use the {\it reduced density matrix} corresponding to each photon $^3$. When the assumption done in references 1 and 2, namely, that each photon in the entangled pair behaves as if it were an individual or isolated photon and obeys Malus' law is abandoned and, as quantum mechanics dictates, the density matrix formalism is used, one can show that the inequalities in question are not violated. 

\

In section 2, we discuss the assumptions involved in the construction of a NLHVT {\it \`a la} Bell $^4$, and derive the Leggett's inequalities at the level of subensambles of entangled pairs of photons with given polarizations.

\

In section 3 we consider two cases for the emitted pair of photons in the zero total angular momentum state: the positive parity state, product of a cascade process in an atom $^5$, and the negative parity state, product of the decay of the ground state of positronium $^6$. Working in the strict framework of quantum mechanics, that is, through the use of the reduced density matrix approach for the calculation of quantum polarization averages of individual but entangled photons, and the well known quantum formulae for the joint detection probabilities $^7$, we show that the Leggett's inequalities are not violated.

\

For completeness, in section 4, we show that the Leggett's inequalities are also not violated by the quantum mechanical description of two entangled spin ${{1}\over {2}}$ particles in the singlet state. 

\

In section 5 we present a conclusion.

\

\

{\bf 2. Leggett's inequalities}

\

Let a source $S$ emit pairs of photons with polarization directions represented by unit vectors $\vec{u}$ and $\vec{v}$, towards corresponding analizers  1 and 2, with orientations given by unit vectors $\vec{a}$ and $\vec{b}$. When each photon of each pair is detected, the results of the polarization measurements are given by functions $A_{\vec{u}}(\vec{a}, \vec{b}; \lambda)$ and $B_{\vec{v}}(\vec{b},\vec{a};\lambda)$ which, at detection, take the values  +1 or -1. (Analizer 1 corresponds to $A$ and analyzer 2 corresponds to $B$.) $\lambda$ is a supplementary (``hidden'') variable taking values in a real domain $\Lambda$, such that for the subensemble of photons with polarizations $\vec{u}$ and $\vec{v}$, has a probability distribution $\rho_{\vec{u},\vec{v}}(\lambda)$ obeying $$\rho_{\vec{u},\vec{v}}(\lambda)\geq 0 \  \  \ and \  \  \ \int_\Lambda d\lambda \rho_{\vec{u},\vec{v}}(\lambda)=1. \eqno{(2.1)}$$ Then one has the following three average values over the subensemble: $$av.(A)=\int_\Lambda d\lambda \rho_{\vec{u},\vec{v}}(\lambda)A_{\vec{u}}(\vec{a},\vec{b};\lambda), \eqno{(2.2)}$$ $$av.(B)=\int_\Lambda d\lambda \rho_{\vec{u},\vec{v}}(\lambda)B_{\vec{v}}(\vec{b},\vec{a};\lambda), \eqno{(2.3)}$$ and $$av.(AB)=\int_\Lambda d\lambda \rho_{\vec{u},\vec{v}}(\lambda)A_{\vec{u}}(\vec{a},\vec{b};\lambda)B_{\vec{v}}(\vec{b},\vec{a};\lambda). \eqno{(2.4)}$$ In principle, $av.(A)$, $av.(B)$ and $av.(AB)$ depend on the set of variables $\{\vec{u},\vec{a},\vec{v},\vec{b}\}$. {\it Non locality} is allowed by the posible dependence of $A$ on $\vec{b}$ and of $B$ on $\vec{a}$, and {\it realism} is represented by the variable $\lambda$ and the assumption that the photons are in the polarization states represented by the vectors $\vec{u}$ and $\vec{v}$.

\
 
In fact, the source emits pairs of photons with polarizations in directions $(\vec{u},\vec{v})$, $(\vec{u}^\prime,\vec{v}^\prime)$, $(\vec{u}^{\prime\prime},\vec{v}^{\prime\prime})$,... (it depends on the decaying state, see next section). However, within each subensemble $(\vec{u}^{(n)},\vec{v}^{(n)})$, the formulae (2.2)-(2.4) for the average values of $A$, $B$, and $AB$ are clearly given in the same spirit of Bell in his original introduction of the classical, unknown, and undetectable hidden variables $\lambda$ in ref. 4. 

\

The quantities $A$ and $B$ obey the conditions listed in the Appendix. Then one has the inequalities $$1-\vert\int_\Lambda d\lambda \rho_{\vec{u},\vec{v}}(\lambda)(A_{\vec{u}}(\vec{a},\vec{b};\lambda)-B_{\vec{v}}(\vec{b},\vec{a};\lambda))\vert \geq \int_\Lambda d\lambda \rho_{\vec{u},\vec{v}}(\lambda)A_{\vec{u}}(\vec{a},\vec{b};\lambda)B_{\vec{v}}(\vec{b},\vec{a};\lambda)$$ $$\geq -1+\vert\int_\Lambda d\lambda \rho_{\vec{u},\vec{v}}(\lambda)(A_{\vec{u}}(\vec{a},\vec{b};\lambda)+B_{\vec{v}}(\vec{b},\vec{a};\lambda))\vert, \eqno{(2.5)}$$ known as Leggett's inequalities. $^1$

\

{\bf 3. Quantum mechanics}

\

We shall now show that quantum mechanics, for certain quantum states, {\it does not} violate the Leggett's inequalities. Coordinates are chosen such that the photons propagate in the $z$-direction, and that polarizations and analyzers are in the $xy$-plane. Photon polarizations are described according to references 7 and 8. We shall consider the following two cases: 

\

(i). Positive parity state: $$|\psi_+>={{1}\over {\sqrt{2}}}(|\vec{x}_1>\otimes|\vec{x}_2>+|\vec{y}_1>\otimes |\vec{y}_2>), \eqno{(3.1)}$$ product of a $J=0\to J=1\to J=0$ radiative cascade decay of calcium $^{5}$. In this case, $\vec{u}=\vec{v}=\vec{x}$ or $\vec{u}=\vec{v}=\vec{y}$.

\

(ii) Negative parity state: $$|\psi_->={{1}\over {\sqrt{2}}}(|\vec{x}_1>\otimes |\vec{y}_2>-|\vec{y}_1>\otimes|\vec{x}_2>), \eqno{(3.2)}$$ product of the ground state  positronium decay $e^+e^-\to \gamma\gamma$. $^{6}$ In this case, $\vec{u}=\vec{x}$ and $\vec{v}=\vec{y}$ or $\vec{u}=\vec{y}$ and $\vec{v}=\vec{x}$.

\

The density operators associated to these states are, respectively, $$\hat{\rho}_+=|\psi_+>\otimes<\psi_+|={{1}\over{2}}(|\vec{x}_1>\otimes|\vec{x}_2>\otimes<\vec{x}_1|\otimes<\vec{x}_2|+|\vec{x}_1>\otimes|\vec{x}_2>\otimes<\vec{y}_1|\otimes<\vec{y}_2|$$ $$+|\vec{y}_1>\otimes|\vec{y}_2>\otimes<\vec{x}_1|\otimes<\vec{x}_2|+|\vec{y}_1>\otimes|\vec{y}_2>\otimes<\vec{y}_1|\otimes<\vec{y}_2|) \eqno{(3.3)}$$ and $$\hat{\rho}_-=|\psi_->\otimes<\psi_-|={{1}\over{2}}(|\vec{x}_1>\otimes|\vec{y}_2>\otimes<\vec{x}_1|\otimes<\vec{y}_2|+|\vec{y}_1>\otimes|\vec{x}_2>\otimes<\vec{y}_1|\otimes<\vec{x}_2|$$ $$-|\vec{y}_1>\otimes|\vec{x}_2>\otimes<\vec{x}_1|\otimes<\vec{y}_2|-|\vec{x}_1>\otimes|\vec{y}_2>\otimes<\vec{y}_1|\otimes<\vec{x}_2|). \eqno{(3.4)}$$ The reduced density operators corresponding to particles $A$ (or 1) ($B$ (or 2)) are obtained by taking the trace over the complete set of states of particle $B$ ($A$). So $$\hat{\rho}^{red.}_{\pm,1}=Tr_2(\hat{\rho}_\pm)=<\vec{x}_2|\hat{\rho}_\pm|\vec{x}_2>+<\vec{y}_2|\hat{\rho}_\pm|\vec{y}_2>={{1}\over{2}}(|\vec{x}_1>\otimes<\vec{x}_1|+|\vec{y}_1>\otimes<\vec{y}_1|), \eqno{(3.5)}$$ $$\hat{\rho}^{red.}_{\pm,2}=Tr_1(\hat{\rho}_\pm)=<\vec{x}_1|\hat{\rho}_\pm|\vec{x}_1>+<\vec{y}_1|\hat{\rho}_\pm|\vec{y}_1>={{1}\over{2}}(|\vec{x}_2>\otimes<\vec{x}_2|+|\vec{y}_2>\otimes<\vec{y}_2|). \eqno{(3.6)}$$ In matrix form, these operators are given by $$\rho^{red.}_{\pm,a}=\pmatrix{<\vec{x}_a|\hat{\rho}^{red.}_{\pm,a}|\vec{x}_a> & <\vec{x}_a|\hat{\rho}^{red.}_{\pm,a}|\vec{y}_a> \cr 
<\vec{y}_a|\hat{\rho}^{red.}_{\pm,a}|\vec{x}_a> & <\vec{y}_a|\hat{\rho}^{red.}_{\pm,a}|\vec{y}_a> \cr}, \ a=1,2$$ and so $$\rho^{red.}_{\pm,1}=\rho^{red.}_{\pm,2}={{1}\over{2}}\pmatrix{1 & 0 \cr 0 & 1}={{1}\over{2}}\I. \eqno{(3.7)}$$ Clearly, this result shows that each photon in the corresponding entangled pair is in a mixed (not pure) state, since $$(\rho^{red.}_{\pm,a})^2={{1}\over{2}}\rho^{red.}_{\pm,a}\neq \rho^{red.}_{\pm,a}, \ a=1,2. \eqno{(3.8)}$$ In other words, each individual photon has no associated wave function. Average values of polarizations must be calculated with the corresponding reduced density matrix operators, and not with a hypotetical individual wave function which does not exist, as was done in references 1 and 2. 

\

The quantum averages corresponding to the hidden variable averages (2.2) and (2.3) are given by the traces of the products of the reduced density operators times the projection operators $$\hat{P}_{\vec{a}}=|\vec{a}>\otimes<\vec{a}|-|\vec{a}^\perp>\otimes<\vec{a}^\perp| \eqno{(3.9)}$$ and $$\hat{P}_{\vec{b}}=|\vec{b}>\otimes<\vec{b}|-|\vec{b}^\perp>\otimes<\vec{b}^\perp| \eqno{(3.10),}$$ where in each case, the first term corresponds to detect the photon with an analizer in direction $\vec{a}$ or $\vec{b}$ (or to go through the ordinary ray in a calcite crystal) and the second term corresponds to the photon to be absorbed (or to go through the extraordinary ray). A straightforward calculation shows that all these averages vanish. Namely: $$av.(A)_{q,+}=Tr_1(\hat{\rho}^{red.}_{+,1}\hat{P}_{\vec{a}})={{1}\over{2}}(|<\vec{a}|\vec{x}_1>|^2-|<\vec{a}^\perp|\vec{x}_1>|^2+|<\vec{a}|\vec{y}_1>|^2-|<\vec{a}^\perp|\vec{y}_1>|^2)=0, \eqno{(3.11)}$$

$$av.(B)_{q,+}=Tr_2(\hat{\rho}^{red.}_{+,2}\hat{P}_{\vec{b}})={{1}\over{2}}(|<\vec{b}|\vec{x}_2>|^2-|<\vec{b}^\perp|\vec{x}_2>|^2+|<\vec{b}|\vec{y}_2>|^2-|<\vec{b}^\perp|\vec{y}_2>|^2)=0, \eqno{(3.12)}$$

$$av.(A)_{q,-}=Tr_1(\hat{\rho}^{red.}_{-,1}\hat{P}_{\vec{a}})={{1}\over{2}}(|<\vec{a}|\vec{x}_1>|^2-|<\vec{a}^\perp|\vec{x}_1>|^2+|<\vec{a}|\vec{y}_1>|^2-|<\vec{a}^\perp|\vec{y}_1>|^2)=0, \eqno{(3.13)}$$

$$av.(B)_{q,-}=Tr_2(\hat{\rho}^{red.}_{-,2}\hat{P}_{\vec{b}})={{1}\over{2}}(|<\vec{b}|\vec{x}_2>|^2-|<\vec{b}^\perp|\vec{x}_2>|^2+|<\vec{b}|\vec{y}_2>|^2-|<\vec{b}^\perp|\vec{y}_2>|^2)=0. \eqno{(3.14)}$$ In each of the four cases, it is easy to verify that the two middle terms cancel to each other, and the same happens between the first and the fourth terms.

\

On the other hand, from the expressions of the well known joint (coincidence) detection quantum probabilities $^{7}$, $$P_{12,q+}(\vec{a},\vec{b})={{1}\over{2}}(\vec{a}\cdot\vec{b})^2 \eqno{(3.15)}$$ and $$P_{12,q-}(\vec{a},\vec{b})={{1}\over{2}}(1-(\vec{a}\cdot\vec{b})^2), \eqno{(3.16)}$$ one easily obtains $$av.(AB)_{q+}=-av.(AB)_{q-}={{1}\over{2}}(2(\vec{a}\cdot \vec{b})^2-1). \eqno{(3.17)}$$ 

\

When the quantum averages (3.11)-(3.14) and (3.17) are replaced in the Leggett's inequalities (2.5), it is immediate to verify that the inequalities are satisfied. In fact, for the case i) one obtains $$1\geq {{1}\over{2}}(2(\vec{a}\cdot\vec{b})^2-1) \geq -1, \eqno{(3.18)}$$ while for the case ii) the result is $$1\geq {{1}\over{2}}(1-2(\vec{a}\cdot\vec{b})^2) \geq -1. \eqno{(3.19)}$$ Both, (3.18) and (3.19), are equivalent to $$1.5\geq(\vec{a}\cdot \vec{b})^2\geq -.5, \eqno{(3.20)}$$ which always holds since $\vec{a}$ and $\vec{b}$ are unit vectors.

\

{\bf 4. Case of spin ${{1}\over{2}}$}

\

For completeness, and with the purpose of giving another example of the fulfillment of the Leggett's inequalities, we consider the case of the EPR-Bohm $^{9,10}$ experiment with two spin ${{1}\over{2}}$ particles $A$ and $B$ (``electrons'') in the singlet state $$|\psi>={{1}\over{\sqrt{2}}}(|\uparrow_A>\otimes |\downarrow_B>-|\downarrow_A>\otimes|\uparrow_B>). \eqno{(4.1)}$$ The density operator is given by $$\hat{\rho}=|\psi>\otimes<\psi|={{1}\over{2}}(|\uparrow_1>\otimes|\downarrow_2>\otimes<\uparrow_1|\otimes<\downarrow_2|-|\uparrow_1>\otimes|\downarrow_2>\otimes<\downarrow_1|\otimes<\uparrow_2|$$ $$-|\downarrow_1>\otimes|\uparrow_2>\otimes<\uparrow_1|\otimes<\downarrow_2|+|\downarrow_1>\otimes|\uparrow_2>\otimes<\downarrow_1|\otimes<\uparrow_2|. \eqno{(4.2)}$$ Then the reduced density operators are $$\hat{\rho}^{red.}_1=Tr_2(\hat{\rho})={{1}\over{2}}(|\downarrow_1>\otimes<\downarrow_1|+|\uparrow_1>\otimes<\uparrow_1|), \eqno{(4.3)}$$ and $$\hat{\rho}^{red.}_2=Tr_1(\hat{\rho})={{1}\over{2}}(|\downarrow_2>\otimes<\downarrow_2|+|\uparrow_2>\otimes<\uparrow_2|). \eqno{(4.4)}$$ In matrix form, these operators are given by $$\hat{\rho}^{red.}_1=\pmatrix{<\uparrow_1|\hat{\rho}^{red.}_1|\uparrow_1> & <\uparrow_1|\hat{\rho}^{red.}_1|\downarrow_1> \cr <\downarrow_1|\hat{\rho}^{red.}_1|\uparrow_1> & <\downarrow_1|\hat{\rho}^{red.}_1|\downarrow_1> \cr}={{1}\over{2}}\pmatrix{1 & 0 \cr 0 & 1 \cr}={{1}\over{2}}\I, \eqno{(4.5)}$$ and $$\hat{\rho}^{red.}_2=\pmatrix{<\uparrow_2|\hat{\rho}^{red.}_2|\uparrow_2> & <\uparrow_2|\hat{\rho}^{red.}_2|\downarrow_2> \cr <\downarrow_2|\hat{\rho}^{red.}_2|\uparrow_2> & <\downarrow_2|\hat{\rho}^{red.}_2|\downarrow_2> \cr}={{1}\over{2}}\pmatrix{1 & 0 \cr 0 & 1 \cr}={{1}\over{2}}\I. \eqno{(4.6)}$$ Again $(\rho_k^{red.})^2={{1}\over{2}}\rho_k^{red.}\neq\rho_k^{red.}$, $k=1,2$, showing that the individual particles are in a mixed (not pure) state; then no individual wave function can be asigned to them. 

\

With projection operators representing Stern-Gerlach apparatuses along unit directions $\vec{a}$ and $\vec{b}$, $$\hat{P}_{\vec{a}}=|\vec{a}>\otimes<\vec{a}|-|-\vec{a}>\otimes<-\vec{a}| \eqno{(4.7)}$$ and  $$\hat{P}_{\vec{b}}=|\vec{b}>\otimes<\vec{b}|-|-\vec{b}>\otimes<-\vec{b}|, \eqno{(4.8)}$$ one obtains the quantum average values $$av.(1)_q(\vec{a})=Tr(\hat{\rho}^{red.}_1\hat{P}_{\vec{a}})={{1}\over{2}}(|<\vec{a}|\uparrow_1>|^2-|<-\vec{a}|\uparrow_1>|^2+|<\vec{a}|\downarrow_1>|^2-|<-\vec{a}|\downarrow_1>|^2)=0 \eqno{(4.9)}$$ and 

$$av.(2)_q(\vec{b})=Tr(\hat{\rho}^{red.}_2\hat{P}_{\vec{b}})={{1}\over{2}}(|<\vec{b}|\uparrow_2>|^2-|<-\vec{b}|\uparrow_2>|^2+|<\vec{b}|\downarrow_2>|^2-|<-\vec{b}|\downarrow_2>|^2)=0. \eqno{(4.10)}$$

\

For the singlet state, the quantum formula for the correlation betwen the two spins is $^4$ $$av.(AB)_q(\uparrow_1,\vec{a};-\downarrow_2,\vec{b})=-\vec{a}\cdot\vec{b}. \eqno{(4.11)}$$ Reinterpreting equation (2.5) for the case of spin ${{1}\over{2}}$ particles, and replacing the hidden variable expectations  by the corresponding quantum formulae, we obtain $$1\geq -\vec{a}\cdot\vec{b}\geq -1, \eqno{(4.12)}$$ which is obviously satisfied.

\

{\bf 5. Conclusion}

\

Both for the case of photon pairs decaying from states with positive and negative parity and zero total angular momentum, as well as for the decay of a singlet state into two spin ${{1}\over{2}}$ particles, we have shown, using the {\it reduced density matrix approach}, that quantum mechanics does {\it not} violate the Leggett's inequalities, derived in the context of NLHVT's. Thus, at the present moment of the development of the subject, the conclusion is that quantum mechanics does not rule out non local realism, and therefore, realism itself. 

\

$(*)$ On sabbatical year at IAFE

\

{\bf Acknowledgements} 

\

This work was partially supported by the project PAPIIT IN113607, DGAPA-UNAM, M\'exico. The author thanks for hospitality at the Facultad de Astronom\'\i a y Astrof\'\i sica de la Universidad de Valencia, Spain, where part of this work was performed, and enlightening discussions with Prof. R. Lapiedra.

\

{\bf References}

\

1. A. J. Leggett, Nonlocal Hidden-Variable Theories and Quantum Mechanics: An Incompatibility Theorem, {\it Foundations of Physics} {\bf 33}, 1469-1493 (2003).

\

2. S. Gr\"oblacher, T. Paterek, R. Kaltenbaek, C. Brukner, M. Zukowski, M. Aspelmayer, and A. Zeilinger, An experimental test of non-local realism, {\it Nature} {\bf 446}, 871-875 (2007).

\

3. C. D. Cantrell, and Marlan O. Scully, The EPR paradox revisited, {\it Physics Reports C} {\bf 43}, 499-508 (1978).

\

4. J. S. Bell, On the Einstein-Podolsky-Rosen paradox, {\it Physics} {\bf 1}, 195-200 (1965).

\

5. A. Aspect, P. Grangier, and G. Rogers, Experimental Tests of Realistic Local Theories via Bell's Theorem, {\it Physical Review Letters} {\bf 47}, 460-463 (1981); A. Peres, {\it Quantum Theory, Concepts and Methods}, Kluwer, The Netherlands (1993).

\

6. L. R. Kasday, J. D. Ullman, and C. S. Wu, Angular Correlation of Compton-Scattered Annihilation Photons and Hidden Variables, {\it Nuovo Cimento B} {\bf 25}, 633-661 (1975); G. Faraci, D. Gutkowski, S. Nottarigo, and A. R. Pennisi, An Experimental Test of the EPR Paradox, {\it Lettere al Nuovo Cimento} {\bf 9}, 607-611 (1974); A. Peres, {\it Quantum Theory, Concepts and Methods}, Kluwer, The Netherlands (1993).

\

7. L. E. Ballentine, {\it Quantum Mechanics}, World Scientific, pp. 597-598 (2001).

\

8. G. Baym, {\it Lectures on Quantum Mechanics}, W. A. Benjamin, New York, pp. 1-8 (1969).

\

9. A. Einstein, B. Podolsky and N. Rosen, Can Quantum-Mechanical Description of Physical Reality Be Considered Complete?, {\it Physical Review} {\bf 47}, 777-780 (1935).

\

10. D. Bohm, {\it Quantum Theory}, Dover, p. 611 (1989).

\

{\bf Appendix}

\

Let $A$ and $B$ be two quantities which take the values +1 and -1. Then it holds $$1-\vert A-B \vert=AB=-1+\vert A+B \vert \eqno{(a.1)}$$ where $\vert \ \ \vert$ denotes the absolute value. Suppose both $A$ and $B$ depend on the variables $\{\lambda, \vec{a}, \vec{b},..., \vec{w} \}$ where $\lambda$ lies in a real domain $\Lambda$ and $\vec{a}$, $\vec{b}$,...,$\vec{w}$ are certain unit vectors in ordinary space. The variable $\lambda$ has a weight given by a classical probability distribution $\rho(\lambda)$ with $\rho(\lambda)\geq 0$ and normalized according to $$\int_\Lambda d\lambda \rho(\lambda)=1. \eqno{(a.2)}$$ From (a.1) and (a.2) $$1-\int_\Lambda d\lambda \rho(\lambda)\vert A-B \vert =\int_\Lambda d\lambda \rho(\lambda) AB =-1+\int_\Lambda d\lambda \rho(\lambda)\vert A+B \vert \eqno{(a.3)}$$ {\it i.e.} $$1-av.(\vert A-B \vert)=av.(AB)=-1+av.(\vert A+B \vert) \eqno{(a.4)}$$ where $$av.(X)=\int_\Lambda d\lambda \rho(\lambda)X(\lambda,\vec{a},\vec{b},...,\vec{w}) \eqno{(a.5)}$$ is the {\it classical average value} of $X$. 

\

Since for any quantity $X$, the average of its absolute value is greater than or equal to the absolute value of its average {\it i.e.} $$av.(\vert X \vert )\geq \vert av.(X) \vert, \eqno{(a.6)}$$ we obtain the inequalities $$1-\vert av.(A)-av.(B)\vert \geq av.(AB) \geq -1+\vert av. (A)+av.(B) \vert \eqno{(a.7)}$$ since $av.(A\pm B)=av.(A)\pm av.(B)$.

\

\

\

\

\

\

e-mails:

socolovs@nucleares.unam.mx, socolovs@iafe.uba.ar

\

\end